\documentclass[aps,prl,twocolumn,superscriptaddress,groupedaddress, floatfix]{revtex4}
\usepackage[utf8]{inputenc}
\usepackage{amsmath}
\usepackage{natbib}
\usepackage{graphicx}
\usepackage[usenames, dvipsnames]{color}
\usepackage{float}
\graphicspath{{Figures/}}
\newcommand{\fA}{f_{\mathrm{A}}}
\newcommand{\gc}{g_{\mathrm{c}}}
\newcommand{\EA}{E_{\mathrm{A}}}
\newcommand{\fr}{f_{\mathrm{r}}}
   
\newcommand{\Vg}{V_{\mathrm{g}}}
\renewcommand{\Im}{I_{\mathrm{m}}}
\newcommand{\Qm}{Q_{\mathrm{m}}}

\newcommand{\Te}{T_{2\mathrm{E}}}

\newcommand{\Tp}{T_\mathrm{parity}}
\newcommand{\ket}[1]{| #1 \rangle}
\newcommand{\bra}[1]{\langle #1 |}
\newcommand{\MHz}{\mathrm{MHz}}
\newcommand{\GHz}{\mathrm{GHz}}

\newcommand{\nm}{\mathrm{nm}}
\newcommand{\mK}{\mathrm{mK}}

\newcommand{\ueV}{\mathrm{\mu}\mathrm{eV}}

\newcommand{\us}{\mathrm{\mu}\mathrm{s}}

\newcommand{\ms}{\mathrm{ms}}
\newcommand{\ns}{\mathrm{ns}}

\newcommand{\g}{\ket{g}}
\newcommand{\e}{\ket{e}}
\newcommand{\Oup}{\ket{o \! \uparrow}}
\newcommand{\Odown}{\ket{o \! \downarrow}}

\begin{document}
\widetext
\title{Direct microwave measurement of Andreev-bound-state dynamics in a proximitized semiconducting nanowire}

\author{M.~Hays}
\email{max.hays@yale.edu}
\affiliation{Department of Applied Physics, Yale University, New Haven, CT 06520, USA}
\author{G.~de~Lange}
\affiliation{Department of Applied Physics, Yale University, New Haven, CT 06520, USA}
\affiliation{QuTech and Delft University of Technology, 2600 GA Delft, The Netherlands}
\affiliation{Kavli Institute of Nanoscience, Delft University of Technology, 2600 GA Delft, The Netherlands}
\author{K.~Serniak}
\affiliation{Department of Applied Physics, Yale University, New Haven, CT 06520, USA}
\author{D.~J.~van Woerkom}
\affiliation{QuTech and Delft University of Technology, 2600 GA Delft, The Netherlands}
\affiliation{Kavli Institute of Nanoscience, Delft University of Technology, 2600 GA Delft, The Netherlands}
\author{D.~Bouman}
\affiliation{QuTech and Delft University of Technology, 2600 GA Delft, The Netherlands}
\affiliation{Kavli Institute of Nanoscience, Delft University of Technology, 2600 GA Delft, The Netherlands}
\author{P.~Krogstrup}
\affiliation{Center for Quantum Devices and Station Q Copenhagen, Niels Bohr Institute,
University of Copenhagen, Universitetsparken 5, 2100 Copenhagen, Denmark}
\author{J.~Nygård}
\affiliation{Center for Quantum Devices and Station Q Copenhagen, Niels Bohr Institute,
University of Copenhagen, Universitetsparken 5, 2100 Copenhagen, Denmark}
\author{A.~Geresdi}
\affiliation{QuTech and Delft University of Technology, 2600 GA Delft, The Netherlands}
\affiliation{Kavli Institute of Nanoscience, Delft University of Technology, 2600 GA Delft, The Netherlands}
\author{M.~H.~Devoret}
\email{michel.devoret@yale.edu}
\affiliation{Department of Applied Physics, Yale University, New Haven, CT 06520, USA}

\date{\today}

\begin{abstract}

The modern understanding of the Josephson effect in mesosopic devices derives from the physics of Andreev bound states, fermionic modes that are localized in a superconducting weak link. 
Recently, Josephson junctions constructed using semiconducting nanowires have led to the realization of superconducting qubits with gate-tunable Josephson energies.
We have used a microwave circuit QED architecture to detect Andreev bound states in such a gate-tunable junction based on an aluminum-proximitized InAs nanowire. 
We demonstrate coherent manipulation of these bound states, and track the bound-state fermion parity in real time.
Individual parity-switching events due to non-equilibrium quasiparticles are observed with a characteristic timescale $\Tp = 160\pm10~\us$. 
The $\Tp$ of a topological nanowire junction sets a lower bound on the bandwidth required for control of Majorana bound states. 
\end{abstract}

\maketitle

The fundamental process governing the physics of mesoscopic superconductors is Andreev reflection, whereby electrons are coherently scattered into holes due to spatial variation of the superconducting order parameter~\cite{andreev}. 
Within a conduction channel of a Josephson junction, constructive interference of Andreev reflection processes results in the formation of localized fermionic modes known as Andreev bound states (ABS). 
These modes have energies less than the superconducting gap, and are responsible for the flow of the Josephson supercurrent~\cite{beenakker1991june,Furusaki1991}. 
While the phenomenological properties of Josephson junctions are widely utilized in superconducting circuits~\cite{clarke2004,devoret2013,Roy2016}, these properties can only be understood in detail by considering the underlying ABS. 

Here we outline the physics of the lowest-energy ABS of a Josephson junction, which is spin-degenerate with energy $\EA$ assuming time-reversal invariance~[Fig.~(1a)].
The many-body configurations of this level can be separated into two manifolds indexed by the parity of fermionic excitations.
The even-parity manifold is spanned by the many-body ground state $\g$ and doubly-excited state $\e$, while the odd-parity manifold is spanned by the singly-excited spin-degenerate states $\Odown$ and $\Oup$.
As the parity-conserving $\g \leftrightarrow \e$ transition involves only discrete sub-gap levels, the even manifold is amenable to coherent manipulation by microwave fields at frequency $\fA = 2\EA/h$~\cite{Zazunov2003, Chtchelkatchev2003, Janvier15}. 
We thus refer to the even manifold as the Andreev qubit.
Dynamics between the even and odd manifolds cannot be controlled, as parity-breaking transitions result from incoherent quasiparticle exchange with the continuum of modes in the environment surrounding the junction~\cite{Zgirski2011,Zazunov2014,Levenson2014}.
However, it is possible to observe these quasiparticle poisoning events by tracking the ABS fermion parity in real time.
The ABS can therefore act as a single-particle detector of the non-equilibrium quasiparticles that plague superconducting devices~\cite{Aumentado,Ferguson,Martinis,Shaw,deVisser,Rist2013}. 
Experiments revealing these dynamics have been performed on ABS hosted by aluminum superconducting atomic contacts~\cite{Janvier15}. 

Advances in the fabrication of superconductor-proximitized semiconducting nanowires~\cite{Chang15,Krogstrup15} have enabled reliable construction of highly-transparent nanowire Josephson junctions (NWJJ).
Due to the low carrier density of semiconductors, the conduction channels of NWJJs can be tuned \textit{in-situ} by electrostatic gates, providing convenient control over the ABS~\cite{VanWoerkom17,Goffman}. 
Such control has been used to create gate-tunable Josephson elements for  superconducting quantum circuits~\cite{deLange15, Larsen15}.
Moreover, high-spin-orbit, large-g-factor NWJJs can in principle be tuned into a topological phase in which the lowest-energy ABS evolves into a Majorana bound state (MBS)~\cite{Lutchyn2010,Oreg2010}. 
As poisoning by non-equilibrium quasiparticles will hinder efforts to probe the physics of MBS~\cite{aasen2016}, monitoring the fermion parity switches of the precursor ABS is a first step towards understanding and mitigating poisoning in a topological NWJJ. 

\begin{figure}
	\centering
	\includegraphics[width=\columnwidth]{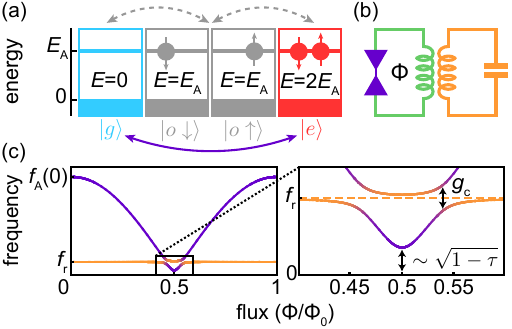} 
	\caption{
    Model of ABS coupled to a microwave resonator. 
	(a)
	Many-body configurations of two spin-degenerate Andreev levels in the excitation representation.  
	Resonant microwaves at frequency $\fA = 2\EA/h$ coherently drive parity-conserving transitions between $\ket{g}$ and $\ket{e}$ (purple arrow). 
	Quasiparticle poisoning induces incoherent transitions between the even- and odd-parity manifolds (gray dashed arrows).
	(b)
	A NWJJ (purple) embedded in a superconducting loop (green). 
    An externally-applied flux $\Phi$ threads the loop, implementing a phase bias $\varphi = 2 \pi \Phi/\Phi_0$ across the junction.
    The loop also realizes an inductive coupling between the NWJJ and a microwave readout resonator (orange, bare frequency $\fr$).
	(c) 
	A representative spectrum of the coupled system consists of $\fA(\Phi)$ (purple) and the resonator transition (orange).
    The maximum of $\fA$ occurs at $\Phi=0$ and depends on the superconducting gap, the junction geometry, and the interface between the superconductor and the NW.
	Zoom: the minimum of $\fA(\Phi)$  occurs at $\Phi=\Phi_0/2$ and is determined by the channel transparency $\tau$. 
	The strength of the inductive coupling is given by $\gc$.
	When the ABS are in the odd manifold they are decoupled from the resonator, leaving only the resonator transition at $\fr$ (dashed orange).	
}
\end{figure}

In this Letter, we report the microwave detection and manipulation of ABS in an aluminum-proximitized indium arsenide (InAs) NWJJ using the techniques of circuit quantum electrodynamics (cQED)~\cite{blais2004,wallraff2004}. 
We perform microwave spectroscopy of a gate- and flux-tunable Andreev qubit, and we achieve coherent manipulation of this Andreev qubit using pulsed microwave fields. 
In addition, we monitor the ABS in real time to directly observe transitions between the even- and odd-parity manifolds, which we attribute to exchange of non-equilibrium quasiparticles between the ABS and the junction leads. 
These parity-switching events are observed to occur with a characteristic timescale $\Tp = 160\pm10~\us$.

Our cQED detection scheme hinges on the supercurrent properties of the ABS. 
While the even manifold supports the flow of supercurrent, the odd manifold does not. 
Therefore, to observe the dynamics of the ABS, we inductively coupled a NWJJ to a superconducting microwave resonator (bare frequency $\fr$) [Fig.~1(b)]~\cite{Janvier15}.
The interaction between the resonator and the current-carrying Andreev qubit is well-described by a conventional Jaynes-Cummings coupling term in the Hamiltonian $\hbar \gc (\hat{a}^\dagger \ket{g}\bra{e}+\hat{a} \ket{e}\bra{g})$, while the current-less odd manifold is decoupled from the resonator [Fig.~1(c)].
When the system is operated in the dispersive regime of cQED such that the magnitude of $\Delta=2\pi(\fA-\fr)$ is much greater than $\gc$, this coupling term takes the form $\hbar \chi \hat{a}^\dagger\hat{a}(\ket{e}\bra{e}-\ket{g}\bra{g})$ where $\chi = \gc^2/\Delta$. 
This results in a qubit-state-dependent shift by $\pm\chi$ of the resonator frequency when the ABS are in the even manifold, while no shift occurs when the ABS are in the odd manifold. 
By monitoring the resonator response to a microwave readout tone, the quantum state of the ABS can be determined. 
However, these frequency shifts must be resolvable with practical measurement integration times.
This requires that $\fA$ be tuned close to $\fr$, which can be achieved by adjusting the superconducting phase difference $\varphi$ and transparency $\tau$ of the conduction channel hosting the lowest-energy ABS doublet [Fig.~1(c)]~\cite{beenakker1991dec}. 
In particular, the conduction channel must be quasi-ballistic
such that $\tau$ can be tuned close to unity~\cite{Janvier15}. 

\begin{figure}
	\centering
	\includegraphics[width=\columnwidth]{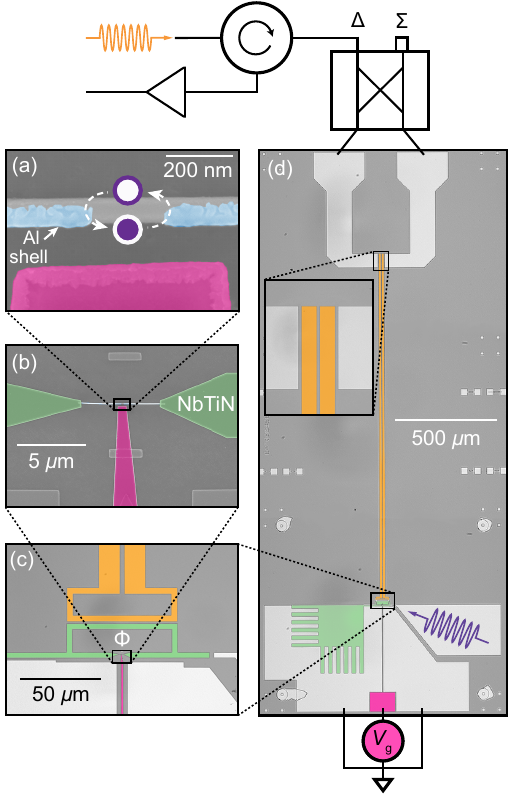} 
	\caption{
    Device and simplified experimental setup.
	(a) 
	Color-enhanced scanning electron micrograph of the InAs NWJJ hosting the ABS. 
    Two of six facets of the NW are coated with a $10~\nm$ thick epitaxial Al shell(blue)~\cite{Krogstrup15}, with a $200~\nm$ gap forming the junction.
	A NbTiN gate (pink) was used for electrostatic tuning of the junction channel transparency $\tau$. 
	(b) 
	Color-enhanced scanning electron micrograph of the Al-coated NW contacted by NbTiN leads (green).
	(c) 
	Color-enhanced optical micrograph of the inductive coupling. 
	An external flux $\Phi$ was applied through a NbTiN loop (green) to phase bias the NWJJ. 
    The loop was inductively coupled to a $\lambda/4$ coplanar stripline resonator (orange), resulting in $\gc/2\pi=23~\MHz$.
	The right side of the loop was capacitively coupled to a microwave drive line (see panel d). 
	(d) 
	Color-enhanced optical micrograph of the full chip. 
    The resonator was measured in reflection using the microwave setup depicted in summary at the top of the figure. 
	A microwave readout tone (orange) with frequency $\fr = 9.066~\GHz$ was transmitted through a $180^{\rm o}$-hybrid, differentially driving the resonator through coupling capacitors (see zoom).
	The reflected microwave tone was routed through a circulator and amplified before being processed at room temperature. 
	On-chip circuitry for electrostatic tuning by $\Vg$ consisted of the gate electrode (pink) and an interdigitated capacitor (green).
    A microwave drive (purple arrow) was used to induce transitions between $\ket{g}$ and $\ket{e}$.
    The smaller features on the sides of the chip are test structures and alignment markers. 
	}
\end{figure}

To achieve a high-$\tau$ NWJJ, we used an MBE-grown [001] wurtzite InAs nanowire [Fig.~2(a)] with an epitaxial aluminum (Al) shell~\cite{Krogstrup15}. 
The device substrate was composed of intrinsic silicon capped with a $300~\nm$ layer of silicon dioxide.
First, the readout resonator ($\fr = 9.066~\GHz$, line width $\kappa/2\pi = 9~\MHz$) and control structures were patterned by electron-beam lithography and reactive ion etching of sputtered niobium titanium nitride (NbTiN).
Then, the nanowire was deposited using a micromanipulator and the junction was defined by selectively wet-etching a 200 nm long section of the Al shell [Fig.~2(a)].
Finally, the Al leads of the NWJJ were contacted to the rest of the circuit with NbTiN [Fig.~2(b)].
We implemented control over $\tau$ via an electrostatic gate voltage $\Vg$ [Fig.~2(a,d)], and we applied an external flux $\Phi$ through a NbTiN loop to bias the NWJJ with phase $\varphi$ [Fig.~2(c)]~\cite{VanWoerkom17}. 
Because the inductance of the NWJJ was much greater than that of the NbTiN loop, $\varphi = 2\pi \Phi/\Phi_0$ where $\Phi_0$ is the magnetic flux quantum. 
A capacitively-coupled microwave drive line was used to drive the $\fA$ transition [Fig.~2(c,d)]. 
In contrast with DC transport measurements, the NWJJ was galvanically isolated from all off-chip circuitry, with an interdigitated capacitor providing an electrostatic reference to the device ground plane [Fig.~2(d)].
The large critical fields of NbTiN and thin-film Al make our devices compatible with high magnetic field measurements, enabling future experiments in the topological regime~\cite{Mourik2012, Krogstrup15}.
The measurements we present here were performed in a dilution refrigerator with a base temperature of $\sim30~\mK$.

\begin{figure}
	\centering
	\includegraphics[width=\columnwidth]{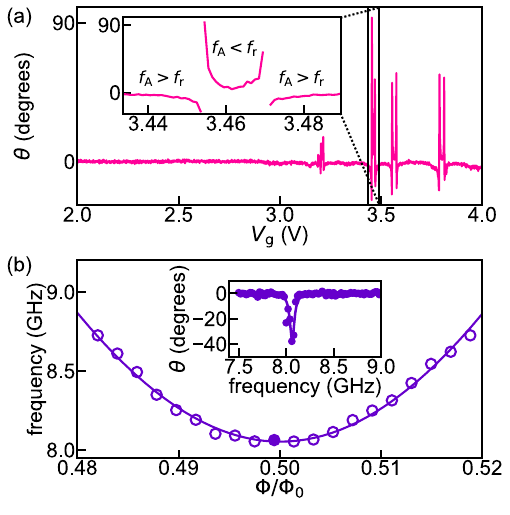} 
	\caption{
	Control of the Andreev qubit frequency. 
    	(a) 
	The average phase $\theta$ of the resonator readout tone for a range of $\Vg$ values ($\Phi=\Phi_0/2$). 
	Each point was integrated for $1.28~\us$. 
	As $\Vg$ is varied, transitions in the nanowire come into proximity with the resonator frequency, resulting in avoided crossings.  
	Inset: zoom on two of these avoided crossings of $\fA$ with the resonator frequency.  
	(b) 
	Inset: Continuous-wave two-tone spectroscopy reveals the qubit transition. 
	The transition frequency $\fA$ is extracted from a best fit to a Lorentzian line shape.
    Main figure: dependence of $\fA$ on $\Phi$.
    Solid line is a fit to the short-junction formula for $\fA$~\cite{beenakker1991dec}.
	}
\end{figure}

We first investigated the effects of $\Vg$ and $\Phi$ on the device properties. 
With the superconducting loop flux-biased to $\Phi_0/2$, we monitored the phase $\theta$ of the resonator readout tone while sweeping $\Vg$ [Fig.~3(a)]. 
For several ranges of $\Vg$, $\theta$ exhibits features consistent with a transition crossing $\fr$ [inset~Fig.~3(a)]. 
We attribute this transition to a gate-controlled Andreev qubit with an inductive coupling to the readout resonator [Fig.~1(b)]. 
The abundance of features observed in Fig.~3(a) may be explained by mesoscopic fluctuations of the nanowire conductance~\cite{ehrenreich1991,VanWoerkom17,Goffman}, with $\fA$ crossing $\fr$ whenever $\tau$ approaches unity [see Fig.~1(c)]. 
Flux-biased two-tone spectroscopy [Fig.~3(b)] performed with $\fA(\Phi_0/2)$ tuned below $\fr$ revealed strong dispersion of $\fA(\Phi)$, consistent with recent observations of ABS in highly transparent InAs/Al NWJJs~\cite{VanWoerkom17}. 
With access only to the low-energy spectrum, a quantitative value for $\tau$ is difficult to obtain. 
However, under the simplifying assumption that $\fA$ is well-described by the short-junction formula $\fA(0) \sqrt{1 - \tau \sin^2(\pi \Phi/\Phi_0)}$~\cite{beenakker1991dec}, we extract $\tau\simeq0.98$ and $\fA(0)\simeq60.0~\GHz  \: (\EA(0) \simeq 124~\ueV)$. 
Drifts in $\Vg$ bias prevented measurements of $\fA$ over a wider flux range.
All measured devices were plagued by these instabilities, which occurred on timescales varying from minutes to hours.   
We attribute these drifts to charging effects in the dielectric surrounding the nanowire. 
While the instabilities made systematic studies requiring long averaging times impossible, they did not inhibit our ability to investigate the fast temporal dynamics of the ABS.

\begin{figure}
	\label{coherent}
	\centering
	\includegraphics[width=\columnwidth]{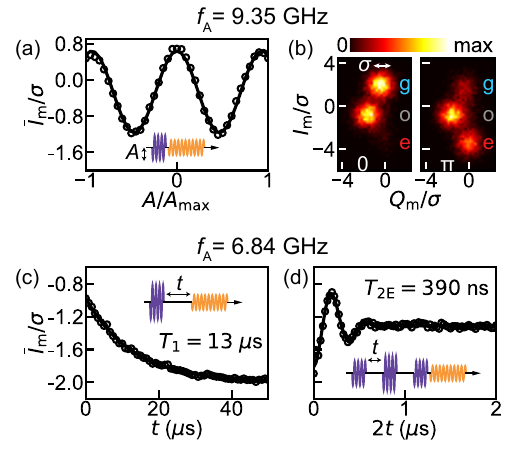} 
	\caption{
	Coherent dynamics of the Andreev qubit.
    The measured values $\Im$, $\Qm$ refer to the in-phase and quadrature components of the reflected readout tone. 
    Here the data are rescaled by the standard deviation $\sigma$ of $\Qm$ when the Andreev qubit is in $\g$. 
	(a) Rabi oscillations of the Andreev qubit  ($\fA = 9.35~\GHz$, $\Phi=\Phi_0/2$). 
	A 10 ns square pulse of varying amplitude $A$ was applied to the qubit, followed by a readout pulse at $\fr$ which was integrated for $640~\ns$.  
	The solid line is a best fit to a sinusoid, used for the calibration of nominal $\pi$ and $\pi/2$ qubit rotations.
	(b) Histogram of the $\Im$ and $\Qm$ quadratures of the resonator readout tone following no qubit rotation (left) and a $\pi$-rotation calibrated from (a) (right).
	(c) Energy relaxation of the qubit ($\fA = 6.84~\GHz$, $\Phi=\Phi_0/2$).  
	Solid line is a best fit to a decaying exponential with time constant $T_1 = 12.8 \pm 0.2 ~\us$.
	(d) Coherence of the qubit measured using a Hahn-echo pulse sequence. 
	The phase of the final $\pi/2$ pulse is varied with the delay to introduce oscillations. 
	Solid line is a best fit to a Gaussian decaying sinusoid with time constant $\Te = 390 \pm 10~\ns$.
	}
\end{figure}

First we probed the coherent dynamics of the $\fA$ transition. 
Fig.~4(a) displays Rabi oscillations of the Andreev qubit at $\Phi = \Phi_0/2$, which were induced by varying the amplitude $~A$ of a $10~\ns$ square pulse with carrier frequency $\fA$.
To verify the effect of the Rabi drive on the ABS, this measurement was performed with high photon number $\bar{n} \sim 100$ and a small detuning $\Delta/2\pi=280~\MHz$.
In this regime, the integrated quadratures $(\Im,~\Qm)$ of the resonator readout pulse clustered into three well-separated Gaussian distributions [Fig.~4(b)]~\cite{Janvier15}. 
We attribute these to $\ket{g}$, $\ket{e}$, and the odd manifold, with the state population indicated by the distribution brightness. 
As expected, the populations of $\g$ and $\e$ change with $A$, while the population of the odd manifold is constant~\cite{SOM}.
The energy and coherence decay of the Andreev qubit were measured at increased detuning to avoid resonator-induced transitions. 
The maximum energy relaxation time $T_1 = 12.8 \pm 0.2 ~\us$ was measured with $\fA=6.84 ~\GHz$ [Fig.~4(c)]. 
At this working point, the Hahn-echo decay time was found to be $\Te = 390 \pm 10 ~\ns$[Fig.~4(d)].
Low-frequency fluctuations in $\fA$ resulted in an immeasurably short Ramsey decay time, which we attribute to the gate-bias instabilities.
The infidelity of the $\pi$-pulse [Fig.~4(b)] is most likely due to this low Ramsey coherence time.
We note that these energy and coherence decay times are of similar magnitude to those observed in Andreev qubits hosted by superconducting atomic contacts ~\cite{Janvier15}, indicating that the loss and dephasing mechanisms at work may be largely independent of the junction materials.  

\begin{figure*}
	\label{quantum_jumps} 
	\centering
	\includegraphics[width=\textwidth]{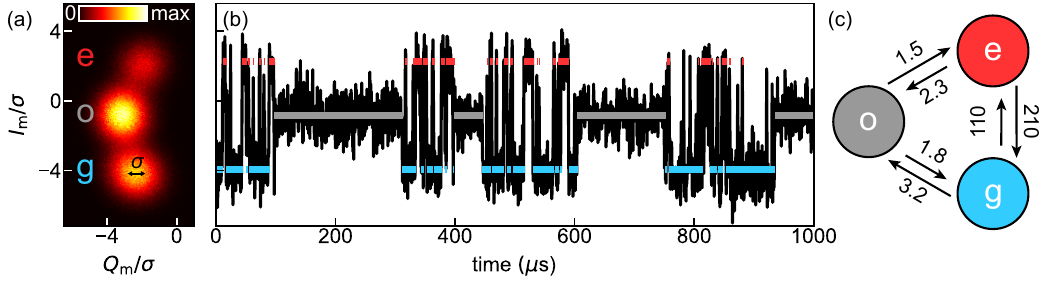} 
	\caption{
	Dynamics of incoherent transitions between many-body configurations of the ABS. 
	(a) 
	Histogram of the $\Im$ and $\Qm$ quadratures of the resonator readout tone ($\fA=8.5~\GHz$, $\Phi = \Phi_0/2$).
	Each count corresponds to an integration period of $480~\ns$ and the total number of counts is $9.6 \times 10^5$. 
	The $(\Im,\Qm)$-pairs cluster into three Gaussian distributions corresponding to the many-body configurations of the ABS [see Fig.~1(a)]. 
	Data are rescaled by the standard deviation $\sigma$ of the $\ket{g}$ distribution.
	(b) Time evolution of $\Im/\sigma$ for a sample of the data in (a). 
	The state assignments shown by the blue, gray, and red bars result from a maximum-likelihood estimation to a hidden Markov model~\cite{SOM}. 
    (c) Transition rates in $\ms^{-1}$ between the ABS many-body configurations extracted from a hidden Markov model~\cite{SOM}. 
	}
\end{figure*}

In addition to studying even-manifold coherence, we observed incoherent transitions between all of the ABS many-body configurations by continuously monitoring the resonator while at small detuning $\Delta/2\pi=-0.5~\GHz$ and high photon number $\bar{n} \sim 100$ [Fig.~5(a,b)]. 
The transition rates between states were extracted by applying a hidden Markov model algorithm to the time evolution of ($\Im$, $\Qm$) [Fig.~(c)]~\cite{press2015, SOM}.
Parity-conserving transitions $\g \leftrightarrow \e$ were off-resonantly driven by the high-power resonator readout tone, resulting in a non-thermal qubit population [Fig.~5(a)] and a reduced qubit lifetime $T_1 = 3.2 \pm 0.1~\us$.
Spontaneous poisoning by non-equilibrium quasiparticles induced parity-breaking transitions between the even and odd manifolds, the rates of which are summarized by the parity-decay timescale $\Tp = 160\pm10~\us$~\cite{SOM}. 
We note that since $\Tp \gg T_1$, the lifetime of the Andreev qubit was limited by direct $\g \leftrightarrow \e$ processes and not by quasiparticle poisoning.  

Previous measurements of bound-state poisoning in proximitized semiconducting nanowires have used Coulomb blockade spectroscopy to estimate the rate of quasiparticle relaxation into a bound state from the proximitizing superconductor~\cite{higginbotham2015,Albrecht2017}. 
Our $\Tp$ measurement is distinct in that we directly monitor the parity of the ABS and are therefore sensitive to all parity-breaking processes. 
To lowest order, the readout tone at $\fr$ should not induce parity-breaking transitions, which involve energies on the order of the superconducting gap. 
However, recent measurements of ABS in superconducting atomic point contacts have shown dependence of $\Tp$ on $\bar{n}$~\cite{JanvierThesis}. 
In future experiments, the dependence of $\Tp$ on $\bar{n}$ will be measured using a Josephson parametric converter~\cite{bergeal2010}.

In conclusion, we have detected and manipulated the ABS of an InAs NWJJ using a cQED approach.
We realized a gate- and flux-tunable Andreev qubit with maximum coherence times $T_1 = 12.8 \pm 0.2 ~\us$ and $\Te = 390 \pm 10 ~\ns$.
Moreover, we achieved continuous monitoring of the ABS fermion parity in a NWJJ, revealing that quasiparticle poisoning of the ABS occurred on a timescale $\Tp = 160\pm10~\us$. 
The measurement time of experiments aiming to detect the non-Abelian properties of MBS in a topological nanowire must fall within a certain range. 
The upper bound on the measurement time is set by $\Tp$, as quasiparticle poisoning of MBS will decohere superpositions of quantum states with different fermion parity. 
Conversely, adiabatic manipulation of MBS restricts the lower bound on the measurement time to nanosecond timescales~\cite{aasen2016}. 
Therefore, our measured value of $\Tp$ leaves an experimentally accessible window for the investigation of Majorana  physics. 

We acknowledge fruitful discussions with Nick Frattini, Sergey Frolov, Luigi Frunzio, Leonid Glazman, Marcelo Goffman, Bernard van Heck, Leo Kouwenhoven, Charlie Marcus, Hugues Pothier, Leandro Tosi, Cristian Urbina, Jukka Väyrynen, Uri Vool, and Shyam Shankar. 
Facilities use was supported by YINQE, the Yale SEAS cleanroom, and NSF MRSEC DMR 1119826. 
This research was supported by ARO under Grant No. W911NF-14-1-0011, by MURI-ONR under Grant No. N00014-16-1-2270, by Microsoft Corporation Station Q, by a Synergy Grant of the European Research Council, and by the Danish National Research Foundation (DG-QDev).
GdL acknowledges support from the European Union’s Horizon 2020 research and innovation programme under the Marie Skłodowska-Curie grant agreement No 656129.
AG acknowledges funding from the Netherlands Organisation for Scientific Research (NWO) through a VENI grant.

\bibliography{biblio}

\begin{thebibliography}{38}
\expandafter\ifx\csname natexlab\endcsname\relax\def\natexlab#1{#1}\fi
\expandafter\ifx\csname bibnamefont\endcsname\relax
  \def\bibnamefont#1{#1}\fi
\expandafter\ifx\csname bibfnamefont\endcsname\relax
  \def\bibfnamefont#1{#1}\fi
\expandafter\ifx\csname citenamefont\endcsname\relax
  \def\citenamefont#1{#1}\fi
\expandafter\ifx\csname url\endcsname\relax
  \def\url#1{\texttt{#1}}\fi
\expandafter\ifx\csname urlprefix\endcsname\relax\def\urlprefix{URL }\fi
\providecommand{\bibinfo}[2]{#2}
\providecommand{\eprint}[2][]{\url{#2}}

\bibitem[{\citenamefont{Andreev}(1964)}]{andreev}
\bibinfo{author}{\bibfnamefont{A.}~\bibnamefont{Andreev}},
  \bibinfo{journal}{Zh. Eksperim. i Teor. Fiz.} \textbf{\bibinfo{volume}{46}}
  (\bibinfo{year}{1964}).

\bibitem[{\citenamefont{Beenakker and Van~Houten}(1991)}]{beenakker1991june}
\bibinfo{author}{\bibfnamefont{C.}~\bibnamefont{Beenakker}} \bibnamefont{and}
  \bibinfo{author}{\bibfnamefont{H.}~\bibnamefont{Van~Houten}},
  \bibinfo{journal}{Phys Rev. Lett.} \textbf{\bibinfo{volume}{66}},
  \bibinfo{pages}{3056} (\bibinfo{year}{1991}).

\bibitem[{\citenamefont{Furusaki and Tsukada}(1991)}]{Furusaki1991}
\bibinfo{author}{\bibfnamefont{A.}~\bibnamefont{Furusaki}} \bibnamefont{and}
  \bibinfo{author}{\bibfnamefont{M.}~\bibnamefont{Tsukada}},
  \bibinfo{journal}{Phys. Rev. B} \textbf{\bibinfo{volume}{43}},
  \bibinfo{pages}{10164} (\bibinfo{year}{1991}).

\bibitem[{\citenamefont{Clarke and Braginski}(2004)}]{clarke2004}
\bibinfo{editor}{\bibfnamefont{J.}~\bibnamefont{Clarke}} \bibnamefont{and}
  \bibinfo{editor}{\bibfnamefont{A.~I.} \bibnamefont{Braginski}}, eds.,
  \emph{\bibinfo{title}{The {SQUID} {Handbook}}}, vol.~\bibinfo{volume}{1}
  (\bibinfo{publisher}{Wiley, Weinheim}, \bibinfo{year}{2004}).

\bibitem[{\citenamefont{Devoret and Schoelkopf}(2013)}]{devoret2013}
\bibinfo{author}{\bibfnamefont{M.~H.} \bibnamefont{Devoret}} \bibnamefont{and}
  \bibinfo{author}{\bibfnamefont{R.~J.} \bibnamefont{Schoelkopf}},
  \bibinfo{journal}{Science} \textbf{\bibinfo{volume}{339}},
  \bibinfo{pages}{1169} (\bibinfo{year}{2013}).

\bibitem[{\citenamefont{Roy and Devoret}(2016)}]{Roy2016}
\bibinfo{author}{\bibfnamefont{A.}~\bibnamefont{Roy}} \bibnamefont{and}
  \bibinfo{author}{\bibfnamefont{M.}~\bibnamefont{Devoret}},
  \bibinfo{journal}{Comptes Rendus Physique} \textbf{\bibinfo{volume}{17}},
  \bibinfo{pages}{740} (\bibinfo{year}{2016}).

\bibitem[{\citenamefont{Zazunov et~al.}(2003)\citenamefont{Zazunov, Shumeiko,
  Bratus, Lantz, and Wendin}}]{Zazunov2003}
\bibinfo{author}{\bibfnamefont{A.}~\bibnamefont{Zazunov}},
  \bibinfo{author}{\bibfnamefont{V.}~\bibnamefont{Shumeiko}},
  \bibinfo{author}{\bibfnamefont{E.}~\bibnamefont{Bratus}},
  \bibinfo{author}{\bibfnamefont{J.}~\bibnamefont{Lantz}}, \bibnamefont{and}
  \bibinfo{author}{\bibfnamefont{G.}~\bibnamefont{Wendin}},
  \bibinfo{journal}{Phys. Rev. Lett.} \textbf{\bibinfo{volume}{90}},
  \bibinfo{pages}{087003} (\bibinfo{year}{2003}).

\bibitem[{\citenamefont{Chtchelkatchev and Nazarov}(2003)}]{Chtchelkatchev2003}
\bibinfo{author}{\bibfnamefont{N.~M.} \bibnamefont{Chtchelkatchev}}
  \bibnamefont{and} \bibinfo{author}{\bibfnamefont{Y.~V.}
  \bibnamefont{Nazarov}}, \bibinfo{journal}{Phys. Rev. Lett.}
  \textbf{\bibinfo{volume}{90}}, \bibinfo{pages}{226806}
  (\bibinfo{year}{2003}).

\bibitem[{\citenamefont{Janvier et~al.}(2015)\citenamefont{Janvier, Tosi,
  Bretheau, Girit, Stern, Bertet, Joyez, Vion, Esteve, Goffman
  et~al.}}]{Janvier15}
\bibinfo{author}{\bibfnamefont{C.}~\bibnamefont{Janvier}},
  \bibinfo{author}{\bibfnamefont{L.}~\bibnamefont{Tosi}},
  \bibinfo{author}{\bibfnamefont{L.}~\bibnamefont{Bretheau}},
  \bibinfo{author}{\bibfnamefont{{\c C}.~{\"O}.} \bibnamefont{Girit}},
  \bibinfo{author}{\bibfnamefont{M.}~\bibnamefont{Stern}},
  \bibinfo{author}{\bibfnamefont{P.}~\bibnamefont{Bertet}},
  \bibinfo{author}{\bibfnamefont{P.}~\bibnamefont{Joyez}},
  \bibinfo{author}{\bibfnamefont{D.}~\bibnamefont{Vion}},
  \bibinfo{author}{\bibfnamefont{D.}~\bibnamefont{Esteve}},
  \bibinfo{author}{\bibfnamefont{M.~F.} \bibnamefont{Goffman}},
  \bibnamefont{et~al.}, \bibinfo{journal}{Science}
  \textbf{\bibinfo{volume}{349}}, \bibinfo{pages}{1199} (\bibinfo{year}{2015}).

\bibitem[{\citenamefont{Zgirski et~al.}(2011)\citenamefont{Zgirski, Bretheau,
  Le~Masne, Pothier, Esteve, and Urbina}}]{Zgirski2011}
\bibinfo{author}{\bibfnamefont{M.}~\bibnamefont{Zgirski}},
  \bibinfo{author}{\bibfnamefont{L.}~\bibnamefont{Bretheau}},
  \bibinfo{author}{\bibfnamefont{Q.}~\bibnamefont{Le~Masne}},
  \bibinfo{author}{\bibfnamefont{H.}~\bibnamefont{Pothier}},
  \bibinfo{author}{\bibfnamefont{D.}~\bibnamefont{Esteve}}, \bibnamefont{and}
  \bibinfo{author}{\bibfnamefont{C.}~\bibnamefont{Urbina}},
  \bibinfo{journal}{Phys. Rev. Lett.} \textbf{\bibinfo{volume}{106}},
  \bibinfo{pages}{257003} (\bibinfo{year}{2011}).

\bibitem[{\citenamefont{Zazunov et~al.}(2014)\citenamefont{Zazunov, Brunetti,
  Yeyati, and Egger}}]{Zazunov2014}
\bibinfo{author}{\bibfnamefont{A.}~\bibnamefont{Zazunov}},
  \bibinfo{author}{\bibfnamefont{A.}~\bibnamefont{Brunetti}},
  \bibinfo{author}{\bibfnamefont{A.~L.} \bibnamefont{Yeyati}},
  \bibnamefont{and} \bibinfo{author}{\bibfnamefont{R.}~\bibnamefont{Egger}},
  \bibinfo{journal}{Phys. Rev. B} \textbf{\bibinfo{volume}{90}},
  \bibinfo{pages}{104508} (\bibinfo{year}{2014}).

\bibitem[{\citenamefont{Levenson-Falk et~al.}(2014)\citenamefont{Levenson-Falk,
  Kos, Vijay, Glazman, and Siddiqi}}]{Levenson2014}
\bibinfo{author}{\bibfnamefont{E.~M.} \bibnamefont{Levenson-Falk}},
  \bibinfo{author}{\bibfnamefont{F.}~\bibnamefont{Kos}},
  \bibinfo{author}{\bibfnamefont{R.}~\bibnamefont{Vijay}},
  \bibinfo{author}{\bibfnamefont{L.}~\bibnamefont{Glazman}}, \bibnamefont{and}
  \bibinfo{author}{\bibfnamefont{I.}~\bibnamefont{Siddiqi}},
  \bibinfo{journal}{Phys. Rev. Lett.} \textbf{\bibinfo{volume}{112}},
  \bibinfo{pages}{047002} (\bibinfo{year}{2014}).

\bibitem[{\citenamefont{Aumentado et~al.}(2004)\citenamefont{Aumentado, Keller,
  Martinis, and Devoret}}]{Aumentado}
\bibinfo{author}{\bibfnamefont{J.}~\bibnamefont{Aumentado}},
  \bibinfo{author}{\bibfnamefont{M.~W.} \bibnamefont{Keller}},
  \bibinfo{author}{\bibfnamefont{J.~M.} \bibnamefont{Martinis}},
  \bibnamefont{and} \bibinfo{author}{\bibfnamefont{M.~H.}
  \bibnamefont{Devoret}}, \bibinfo{journal}{Phys. Rev. Lett.}
  \textbf{\bibinfo{volume}{92}}, \bibinfo{pages}{066802}
  (\bibinfo{year}{2004}).

\bibitem[{\citenamefont{Ferguson et~al.}(2006)\citenamefont{Ferguson, Andresen,
  Brenner, and Clark}}]{Ferguson}
\bibinfo{author}{\bibfnamefont{A.~J.} \bibnamefont{Ferguson}},
  \bibinfo{author}{\bibfnamefont{S.~E.} \bibnamefont{Andresen}},
  \bibinfo{author}{\bibfnamefont{R.}~\bibnamefont{Brenner}}, \bibnamefont{and}
  \bibinfo{author}{\bibfnamefont{R.~G.} \bibnamefont{Clark}},
  \bibinfo{journal}{Phys. Rev. Lett.} \textbf{\bibinfo{volume}{97}},
  \bibinfo{pages}{086602} (\bibinfo{year}{2006}).

\bibitem[{\citenamefont{Martinis et~al.}(2009)\citenamefont{Martinis, Ansmann,
  and Aumentado}}]{Martinis}
\bibinfo{author}{\bibfnamefont{J.~M.} \bibnamefont{Martinis}},
  \bibinfo{author}{\bibfnamefont{M.}~\bibnamefont{Ansmann}}, \bibnamefont{and}
  \bibinfo{author}{\bibfnamefont{J.}~\bibnamefont{Aumentado}},
  \bibinfo{journal}{Phys. Rev. Lett.} \textbf{\bibinfo{volume}{103}},
  \bibinfo{pages}{097002} (\bibinfo{year}{2009}).

\bibitem[{\citenamefont{Shaw et~al.}(2008)\citenamefont{Shaw, Lutchyn, Delsing,
  and Echternach}}]{Shaw}
\bibinfo{author}{\bibfnamefont{M.~D.} \bibnamefont{Shaw}},
  \bibinfo{author}{\bibfnamefont{R.~M.} \bibnamefont{Lutchyn}},
  \bibinfo{author}{\bibfnamefont{P.}~\bibnamefont{Delsing}}, \bibnamefont{and}
  \bibinfo{author}{\bibfnamefont{P.~M.} \bibnamefont{Echternach}},
  \bibinfo{journal}{Phys. Rev. B} \textbf{\bibinfo{volume}{78}},
  \bibinfo{pages}{024503} (\bibinfo{year}{2008}).

\bibitem[{\citenamefont{de~Visser et~al.}(2011)\citenamefont{de~Visser,
  Baselmans, Diener, Yates, Endo, and Klapwijk}}]{deVisser}
\bibinfo{author}{\bibfnamefont{P.~J.} \bibnamefont{de~Visser}},
  \bibinfo{author}{\bibfnamefont{J.~J.~A.} \bibnamefont{Baselmans}},
  \bibinfo{author}{\bibfnamefont{P.}~\bibnamefont{Diener}},
  \bibinfo{author}{\bibfnamefont{S.~J.~C.} \bibnamefont{Yates}},
  \bibinfo{author}{\bibfnamefont{A.}~\bibnamefont{Endo}}, \bibnamefont{and}
  \bibinfo{author}{\bibfnamefont{T.~M.} \bibnamefont{Klapwijk}},
  \bibinfo{journal}{Phys. Rev. Lett.} \textbf{\bibinfo{volume}{106}},
  \bibinfo{pages}{167004} (\bibinfo{year}{2011}).

\bibitem[{\citenamefont{Rist{\`{e}} et~al.}(2013)\citenamefont{Rist{\`{e}},
  Bultink, Tiggelman, Schouten, Lehnert, and DiCarlo}}]{Rist2013}
\bibinfo{author}{\bibfnamefont{D.}~\bibnamefont{Rist{\`{e}}}},
  \bibinfo{author}{\bibfnamefont{C.~C.} \bibnamefont{Bultink}},
  \bibinfo{author}{\bibfnamefont{M.~J.} \bibnamefont{Tiggelman}},
  \bibinfo{author}{\bibfnamefont{R.~N.} \bibnamefont{Schouten}},
  \bibinfo{author}{\bibfnamefont{K.~W.} \bibnamefont{Lehnert}},
  \bibnamefont{and} \bibinfo{author}{\bibfnamefont{L.}~\bibnamefont{DiCarlo}},
  \bibinfo{journal}{Nature Communications} \textbf{\bibinfo{volume}{4}},
  \bibinfo{pages}{1913} (\bibinfo{year}{2013}).

\bibitem[{\citenamefont{Chang et~al.}(2015)\citenamefont{Chang, Albrecht,
  Jespersen, Kuemmeth, Krogstrup, Nyg{\aa}rd, and Marcus}}]{Chang15}
\bibinfo{author}{\bibfnamefont{W.}~\bibnamefont{Chang}},
  \bibinfo{author}{\bibfnamefont{S.}~\bibnamefont{Albrecht}},
  \bibinfo{author}{\bibfnamefont{T.}~\bibnamefont{Jespersen}},
  \bibinfo{author}{\bibfnamefont{F.}~\bibnamefont{Kuemmeth}},
  \bibinfo{author}{\bibfnamefont{P.}~\bibnamefont{Krogstrup}},
  \bibinfo{author}{\bibfnamefont{J.}~\bibnamefont{Nyg{\aa}rd}},
  \bibnamefont{and} \bibinfo{author}{\bibfnamefont{C.}~\bibnamefont{Marcus}},
  \bibinfo{journal}{Nat. Nanotechnol.} \textbf{\bibinfo{volume}{10}},
  \bibinfo{pages}{232} (\bibinfo{year}{2015}).

\bibitem[{\citenamefont{Krogstrup et~al.}(2015)\citenamefont{Krogstrup, Ziino,
  Chang, Albrecht, Madsen, Johnson, Nyg{\aa}rd, Marcus, and
  Jespersen}}]{Krogstrup15}
\bibinfo{author}{\bibfnamefont{P.}~\bibnamefont{Krogstrup}},
  \bibinfo{author}{\bibfnamefont{N.~L.~B.} \bibnamefont{Ziino}},
  \bibinfo{author}{\bibfnamefont{W.}~\bibnamefont{Chang}},
  \bibinfo{author}{\bibfnamefont{S.~M.} \bibnamefont{Albrecht}},
  \bibinfo{author}{\bibfnamefont{M.~H.} \bibnamefont{Madsen}},
  \bibinfo{author}{\bibfnamefont{E.}~\bibnamefont{Johnson}},
  \bibinfo{author}{\bibfnamefont{J.}~\bibnamefont{Nyg{\aa}rd}},
  \bibinfo{author}{\bibfnamefont{C.~M.} \bibnamefont{Marcus}},
  \bibnamefont{and} \bibinfo{author}{\bibfnamefont{T.~S.}
  \bibnamefont{Jespersen}}, \bibinfo{journal}{Nat. Mater.}
  \textbf{\bibinfo{volume}{14}}, \bibinfo{pages}{400} (\bibinfo{year}{2015}).

\bibitem[{\citenamefont{van Woerkom et~al.}(2017)\citenamefont{van Woerkom,
  Proutski, van Heck, Bouman, V{\"a}yrynen, Glazman, Krogstrup, Nyg{\aa}rd,
  Kouwenhoven, and Geresdi}}]{VanWoerkom17}
\bibinfo{author}{\bibfnamefont{D.~J.} \bibnamefont{van Woerkom}},
  \bibinfo{author}{\bibfnamefont{A.}~\bibnamefont{Proutski}},
  \bibinfo{author}{\bibfnamefont{B.}~\bibnamefont{van Heck}},
  \bibinfo{author}{\bibfnamefont{D.}~\bibnamefont{Bouman}},
  \bibinfo{author}{\bibfnamefont{J.~I.} \bibnamefont{V{\"a}yrynen}},
  \bibinfo{author}{\bibfnamefont{L.~I.} \bibnamefont{Glazman}},
  \bibinfo{author}{\bibfnamefont{P.}~\bibnamefont{Krogstrup}},
  \bibinfo{author}{\bibfnamefont{J.}~\bibnamefont{Nyg{\aa}rd}},
  \bibinfo{author}{\bibfnamefont{L.~P.} \bibnamefont{Kouwenhoven}},
  \bibnamefont{and} \bibinfo{author}{\bibfnamefont{A.}~\bibnamefont{Geresdi}},
  \bibinfo{journal}{Nature Phys.}  (\bibinfo{year}{2017}).

\bibitem[{\citenamefont{Goffman et~al.}(2017)\citenamefont{Goffman, Urbina,
  Pothier, Nygård, Marcus, and Krogstrup}}]{Goffman}
\bibinfo{author}{\bibfnamefont{M.~F.} \bibnamefont{Goffman}},
  \bibinfo{author}{\bibfnamefont{C.}~\bibnamefont{Urbina}},
  \bibinfo{author}{\bibfnamefont{H.}~\bibnamefont{Pothier}},
  \bibinfo{author}{\bibfnamefont{J.}~\bibnamefont{Nygård}},
  \bibinfo{author}{\bibfnamefont{C.~M.} \bibnamefont{Marcus}},
  \bibnamefont{and}
  \bibinfo{author}{\bibfnamefont{P.}~\bibnamefont{Krogstrup}},
  \bibinfo{journal}{New Journal of Physics} \textbf{\bibinfo{volume}{19}},
  \bibinfo{pages}{092002} (\bibinfo{year}{2017}).

\bibitem[{\citenamefont{de~Lange et~al.}(2015)\citenamefont{de~Lange, van Heck,
  Bruno, van Woerkom, Geresdi, Plissard, Bakkers, Akhmerov, and
  DiCarlo}}]{deLange15}
\bibinfo{author}{\bibfnamefont{G.}~\bibnamefont{de~Lange}},
  \bibinfo{author}{\bibfnamefont{B.}~\bibnamefont{van Heck}},
  \bibinfo{author}{\bibfnamefont{A.}~\bibnamefont{Bruno}},
  \bibinfo{author}{\bibfnamefont{D.~J.} \bibnamefont{van Woerkom}},
  \bibinfo{author}{\bibfnamefont{A.}~\bibnamefont{Geresdi}},
  \bibinfo{author}{\bibfnamefont{S.~R.} \bibnamefont{Plissard}},
  \bibinfo{author}{\bibfnamefont{E.~P. A.~M.} \bibnamefont{Bakkers}},
  \bibinfo{author}{\bibfnamefont{A.~R.} \bibnamefont{Akhmerov}},
  \bibnamefont{and} \bibinfo{author}{\bibfnamefont{L.}~\bibnamefont{DiCarlo}},
  \bibinfo{journal}{Phys. Rev. Lett.} \textbf{\bibinfo{volume}{115}},
  \bibinfo{pages}{127002} (\bibinfo{year}{2015}).

\bibitem[{\citenamefont{Larsen et~al.}(2015)\citenamefont{Larsen, Petersson,
  Kuemmeth, Jespersen, Krogstrup, Nyg\aa{}rd, and Marcus}}]{Larsen15}
\bibinfo{author}{\bibfnamefont{T.~W.} \bibnamefont{Larsen}},
  \bibinfo{author}{\bibfnamefont{K.~D.} \bibnamefont{Petersson}},
  \bibinfo{author}{\bibfnamefont{F.}~\bibnamefont{Kuemmeth}},
  \bibinfo{author}{\bibfnamefont{T.~S.} \bibnamefont{Jespersen}},
  \bibinfo{author}{\bibfnamefont{P.}~\bibnamefont{Krogstrup}},
  \bibinfo{author}{\bibfnamefont{J.}~\bibnamefont{Nyg\aa{}rd}},
  \bibnamefont{and} \bibinfo{author}{\bibfnamefont{C.~M.}
  \bibnamefont{Marcus}}, \bibinfo{journal}{Phys. Rev. Lett.}
  \textbf{\bibinfo{volume}{115}}, \bibinfo{pages}{127001}
  (\bibinfo{year}{2015}).

\bibitem[{\citenamefont{Lutchyn et~al.}(2010)\citenamefont{Lutchyn, Sau, and
  Das~Sarma}}]{Lutchyn2010}
\bibinfo{author}{\bibfnamefont{R.~M.} \bibnamefont{Lutchyn}},
  \bibinfo{author}{\bibfnamefont{J.~D.} \bibnamefont{Sau}}, \bibnamefont{and}
  \bibinfo{author}{\bibfnamefont{S.}~\bibnamefont{Das~Sarma}},
  \bibinfo{journal}{Phys. Rev. Lett.} \textbf{\bibinfo{volume}{105}},
  \bibinfo{pages}{077001} (\bibinfo{year}{2010}).

\bibitem[{\citenamefont{Oreg et~al.}(2010)\citenamefont{Oreg, Refael, and von
  Oppen}}]{Oreg2010}
\bibinfo{author}{\bibfnamefont{Y.}~\bibnamefont{Oreg}},
  \bibinfo{author}{\bibfnamefont{G.}~\bibnamefont{Refael}}, \bibnamefont{and}
  \bibinfo{author}{\bibfnamefont{F.}~\bibnamefont{von Oppen}},
  \bibinfo{journal}{Phys. Rev. Lett.} \textbf{\bibinfo{volume}{105}},
  \bibinfo{pages}{177002} (\bibinfo{year}{2010}).

\bibitem[{\citenamefont{Aasen et~al.}(2016)\citenamefont{Aasen, Hell, Mishmash,
  Higginbotham, Danon, Leijnse, Jespersen, Folk, Marcus, Flensberg
  et~al.}}]{aasen2016}
\bibinfo{author}{\bibfnamefont{D.}~\bibnamefont{Aasen}},
  \bibinfo{author}{\bibfnamefont{M.}~\bibnamefont{Hell}},
  \bibinfo{author}{\bibfnamefont{R.~V.} \bibnamefont{Mishmash}},
  \bibinfo{author}{\bibfnamefont{A.}~\bibnamefont{Higginbotham}},
  \bibinfo{author}{\bibfnamefont{J.}~\bibnamefont{Danon}},
  \bibinfo{author}{\bibfnamefont{M.}~\bibnamefont{Leijnse}},
  \bibinfo{author}{\bibfnamefont{T.~S.} \bibnamefont{Jespersen}},
  \bibinfo{author}{\bibfnamefont{J.~A.} \bibnamefont{Folk}},
  \bibinfo{author}{\bibfnamefont{C.~M.} \bibnamefont{Marcus}},
  \bibinfo{author}{\bibfnamefont{K.}~\bibnamefont{Flensberg}},
  \bibnamefont{et~al.}, \bibinfo{journal}{Phys. Rev. X}
  \textbf{\bibinfo{volume}{6}}, \bibinfo{pages}{031016} (\bibinfo{year}{2016}).

\bibitem[{\citenamefont{Blais et~al.}(2004)\citenamefont{Blais, Huang,
  Wallraff, Girvin, and Schoelkopf}}]{blais2004}
\bibinfo{author}{\bibfnamefont{A.}~\bibnamefont{Blais}},
  \bibinfo{author}{\bibfnamefont{R.-S.} \bibnamefont{Huang}},
  \bibinfo{author}{\bibfnamefont{A.}~\bibnamefont{Wallraff}},
  \bibinfo{author}{\bibfnamefont{S.~M.} \bibnamefont{Girvin}},
  \bibnamefont{and} \bibinfo{author}{\bibfnamefont{R.~J.}
  \bibnamefont{Schoelkopf}}, \bibinfo{journal}{Phys. Rev. A}
  \textbf{\bibinfo{volume}{69}}, \bibinfo{pages}{062320}
  (\bibinfo{year}{2004}).

\bibitem[{\citenamefont{Wallraff et~al.}(2004)\citenamefont{Wallraff, Schuster,
  Blais, Frunzio et~al.}}]{wallraff2004}
\bibinfo{author}{\bibfnamefont{A.}~\bibnamefont{Wallraff}},
  \bibinfo{author}{\bibfnamefont{D.~I.} \bibnamefont{Schuster}},
  \bibinfo{author}{\bibfnamefont{A.}~\bibnamefont{Blais}},
  \bibinfo{author}{\bibfnamefont{L.}~\bibnamefont{Frunzio}},
  \bibnamefont{et~al.}, \bibinfo{journal}{Nature}
  \textbf{\bibinfo{volume}{431}}, \bibinfo{pages}{162} (\bibinfo{year}{2004}).

\bibitem[{\citenamefont{Beenakker}(1991)}]{beenakker1991dec}
\bibinfo{author}{\bibfnamefont{C.~W.~J.} \bibnamefont{Beenakker}},
  \bibinfo{journal}{Phys. Rev. Lett.} \textbf{\bibinfo{volume}{67}},
  \bibinfo{pages}{3836} (\bibinfo{year}{1991}).

\bibitem[{\citenamefont{Mourik et~al.}(2012)\citenamefont{Mourik, Zuo, Frolov,
  Plissard, Bakkers, and Kouwenhoven}}]{Mourik2012}
\bibinfo{author}{\bibfnamefont{V.}~\bibnamefont{Mourik}},
  \bibinfo{author}{\bibfnamefont{K.}~\bibnamefont{Zuo}},
  \bibinfo{author}{\bibfnamefont{S.~M.} \bibnamefont{Frolov}},
  \bibinfo{author}{\bibfnamefont{S.~R.} \bibnamefont{Plissard}},
  \bibinfo{author}{\bibfnamefont{E.~P. A.~M.} \bibnamefont{Bakkers}},
  \bibnamefont{and} \bibinfo{author}{\bibfnamefont{L.~P.}
  \bibnamefont{Kouwenhoven}}, \bibinfo{journal}{Science}
  \textbf{\bibinfo{volume}{336}}, \bibinfo{pages}{1003} (\bibinfo{year}{2012}).

\bibitem[{\citenamefont{Ehrenreich and Turnbull}(1991)}]{ehrenreich1991}
\bibinfo{author}{\bibfnamefont{H.}~\bibnamefont{Ehrenreich}} \bibnamefont{and}
  \bibinfo{author}{\bibfnamefont{D.}~\bibnamefont{Turnbull}},
  \emph{\bibinfo{title}{Advances in Research and Applications: Semiconductor
  Heterostructures and Nanostructures}}, vol.~\bibinfo{volume}{44}
  (\bibinfo{publisher}{Academic Press}, \bibinfo{year}{1991}).

\bibitem[{SOM()}]{SOM}
\emph{\bibinfo{title}{See supplemental material}}.

\bibitem[{\citenamefont{Press et~al.}(2015)\citenamefont{Press, Teukolsky,
  Vetterling, and Flannery}}]{press2015}
\bibinfo{author}{\bibfnamefont{W.~H.} \bibnamefont{Press}},
  \bibinfo{author}{\bibfnamefont{S.~A.} \bibnamefont{Teukolsky}},
  \bibinfo{author}{\bibfnamefont{W.~T.} \bibnamefont{Vetterling}},
  \bibnamefont{and} \bibinfo{author}{\bibfnamefont{B.~P.}
  \bibnamefont{Flannery}}, \emph{\bibinfo{title}{Numerical recipes in C++}}
  (\bibinfo{year}{2015}).

\bibitem[{\citenamefont{Higginbotham et~al.}(2015)\citenamefont{Higginbotham,
  Albrecht, Kir{\v{s}}anskas, Chang, Kuemmeth, Krogstrup, Jespersen,
  Nyg{\aa}rd, Flensberg, and Marcus}}]{higginbotham2015}
\bibinfo{author}{\bibfnamefont{A.~P.} \bibnamefont{Higginbotham}},
  \bibinfo{author}{\bibfnamefont{S.~M.} \bibnamefont{Albrecht}},
  \bibinfo{author}{\bibfnamefont{G.}~\bibnamefont{Kir{\v{s}}anskas}},
  \bibinfo{author}{\bibfnamefont{W.}~\bibnamefont{Chang}},
  \bibinfo{author}{\bibfnamefont{F.}~\bibnamefont{Kuemmeth}},
  \bibinfo{author}{\bibfnamefont{P.}~\bibnamefont{Krogstrup}},
  \bibinfo{author}{\bibfnamefont{T.~S.} \bibnamefont{Jespersen}},
  \bibinfo{author}{\bibfnamefont{J.}~\bibnamefont{Nyg{\aa}rd}},
  \bibinfo{author}{\bibfnamefont{K.}~\bibnamefont{Flensberg}},
  \bibnamefont{and} \bibinfo{author}{\bibfnamefont{C.~M.}
  \bibnamefont{Marcus}}, \bibinfo{journal}{Nature Phys.}
  \textbf{\bibinfo{volume}{11}}, \bibinfo{pages}{1017} (\bibinfo{year}{2015}).

\bibitem[{\citenamefont{Albrecht et~al.}(2017)\citenamefont{Albrecht, Hansen,
  Higginbotham, Kuemmeth, Jespersen, Nyg{\aa}rd, Krogstrup, Danon, Flensberg,
  and Marcus}}]{Albrecht2017}
\bibinfo{author}{\bibfnamefont{S.}~\bibnamefont{Albrecht}},
  \bibinfo{author}{\bibfnamefont{E.}~\bibnamefont{Hansen}},
  \bibinfo{author}{\bibfnamefont{A.}~\bibnamefont{Higginbotham}},
  \bibinfo{author}{\bibfnamefont{F.}~\bibnamefont{Kuemmeth}},
  \bibinfo{author}{\bibfnamefont{T.}~\bibnamefont{Jespersen}},
  \bibinfo{author}{\bibfnamefont{J.}~\bibnamefont{Nyg{\aa}rd}},
  \bibinfo{author}{\bibfnamefont{P.}~\bibnamefont{Krogstrup}},
  \bibinfo{author}{\bibfnamefont{J.}~\bibnamefont{Danon}},
  \bibinfo{author}{\bibfnamefont{K.}~\bibnamefont{Flensberg}},
  \bibnamefont{and} \bibinfo{author}{\bibfnamefont{C.}~\bibnamefont{Marcus}},
  \bibinfo{journal}{Phys. Rev. Lett.} \textbf{\bibinfo{volume}{118}},
  \bibinfo{pages}{137701} (\bibinfo{year}{2017}).

\bibitem[{\citenamefont{Janvier}(2016)}]{JanvierThesis}
\bibinfo{author}{\bibfnamefont{C.}~\bibnamefont{Janvier}},
  \bibinfo{type}{Thesis}, \bibinfo{school}{{Universit{\'e} Paris-Saclay}}
  (\bibinfo{year}{2016}).

\bibitem[{\citenamefont{Bergeal et~al.}(2010)\citenamefont{Bergeal, Schackert,
  Metcalfe, Vijay, Manucharyan, Frunzio, Prober, Schoelkopf, Girvin, and
  Devoret}}]{bergeal2010}
\bibinfo{author}{\bibfnamefont{N.}~\bibnamefont{Bergeal}},
  \bibinfo{author}{\bibfnamefont{F.}~\bibnamefont{Schackert}},
  \bibinfo{author}{\bibfnamefont{M.}~\bibnamefont{Metcalfe}},
  \bibinfo{author}{\bibfnamefont{R.}~\bibnamefont{Vijay}},
  \bibinfo{author}{\bibfnamefont{V.~E.} \bibnamefont{Manucharyan}},
  \bibinfo{author}{\bibfnamefont{L.}~\bibnamefont{Frunzio}},
  \bibinfo{author}{\bibfnamefont{D.~E.} \bibnamefont{Prober}},
  \bibinfo{author}{\bibfnamefont{R.~J.} \bibnamefont{Schoelkopf}},
  \bibinfo{author}{\bibfnamefont{S.~M.} \bibnamefont{Girvin}},
  \bibnamefont{and} \bibinfo{author}{\bibfnamefont{M.~H.}
  \bibnamefont{Devoret}}, \bibinfo{journal}{Nature}
  \textbf{\bibinfo{volume}{465}}, \bibinfo{pages}{64} (\bibinfo{year}{2010}).

\end{thebibliography}


\begin{thebibliography}{11}
\bibitem{purcell} E. M. Purcell, in Confined Electrons and Photons (Springer, 1995), pp. 839-839.
\bibitem{press} W. H. Press, S. A. Teukolsky, W. T. Vetterling, and B. P.
Flannery, Numerical recipes in C++ (2015).
\bibitem{vool2014} U. Vool, I. M. Pop, K. Sliwa, B. Abdo, C. Wang,T. Brecht, Y. Y. Gao, S. Shankar, M. Hatridge, G. Catelani, et al., Physical review letters 113, 247001 (2014).
\end{thebibliography}

\clearpage
\widetext
\begin{center}
\textbf{\large Supplemental materials for ``Direct microwave measurement of Andreev-bound-state dynamics in a proximitized semiconducting nanowire''}
\end{center}

\setcounter{equation}{0}
\setcounter{figure}{0}
\setcounter{table}{0}
\setcounter{page}{1}
\makeatletter
\renewcommand{\theequation}{S\arabic{equation}}
\renewcommand{\thefigure}{S\arabic{figure}}
\renewcommand{\bibnumfmt}[1]{[S#1]}
\renewcommand{\citenumfont}[1]{S#1}

\section{EFFECT OF RABI DRIVE ON ABS MANY-BODY CONFIGURATIONS}

As discussed in the main text, it was observed that the Rabi drive on the Andreev qubit did not affect the population of the odd manifold [Fig. 4(b)].
To see this quantitatively, we projected the histograms of Fig. 4(b) onto the $\Im$-axis [Fig. S1]. 
By fitting to Gaussian distributions, the population of $\g$, $\e$, and the odd manifold were estimated. 
The fits yield that the equilibrium population of the odd manifold is 
$0.49 \pm 0.01$, while the population of the odd manifold following the $\pi$ pulse is $0.52\pm0.02$. 
We thus observe that the odd manifold population is unaffected by the $\pi$ pulse within uncertainty. 
The residual $\g$ population following the $\pi$-pulse is most likely due to fluctuations of the Andreev qubit transition frequency as well as qubit relaxation events due to the Purcell effect \cite{purcell}. 

\begin{figure}[h]
	\centering
	\includegraphics[width=5in]{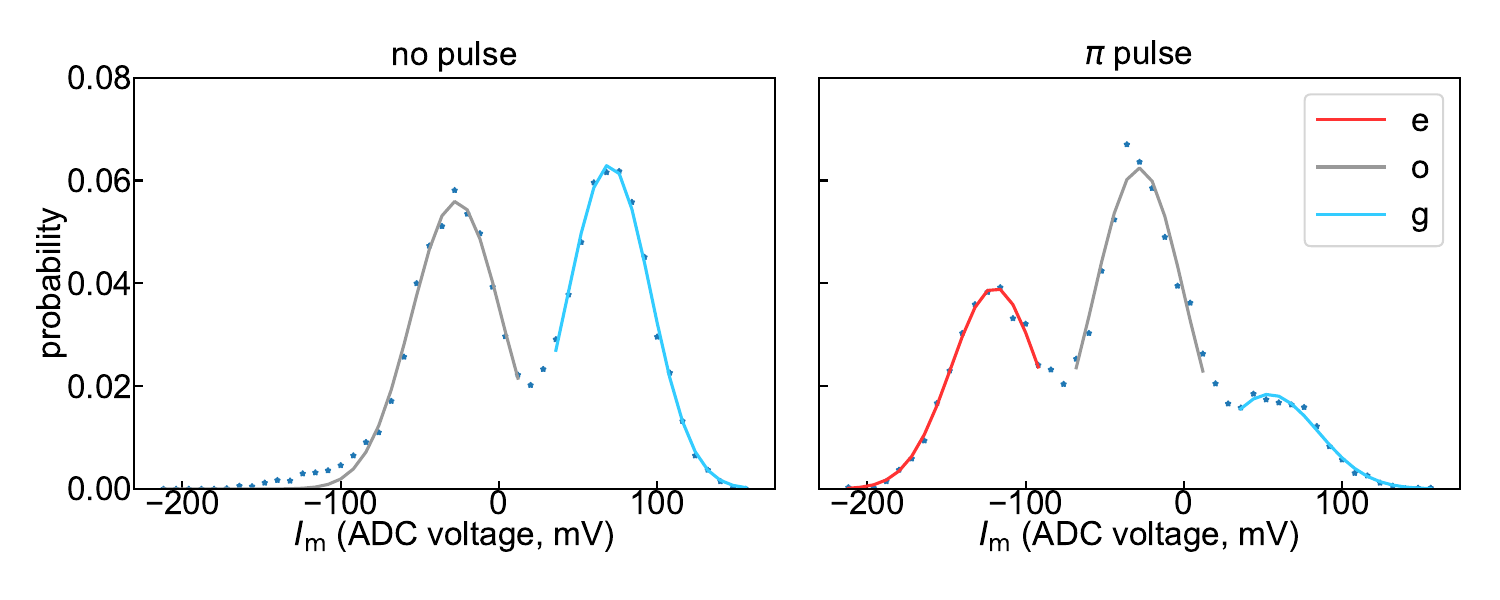} 
	\caption{
    Histograms of Fig. 4(b) projected onto the $\Im$-axis. 
    The colored lines are Gaussian fits. 
    The left panel shows the equilibrium $\Im$ distribution, while the right panel displays the $\Im$ distribution following a $\pi$ pulse. 
	}
    \label{single_shot}
\end{figure}

\section{QUANTUM JUMP ANALYSIS}

We now discuss in further detail the analysis of the time evolution of ($\Im$, $\Qm$) displayed in Fig. 5(b). 
Following the hidden Markov algorithm outlined in \cite{press}, we extracted the transition rates $\Gamma_{ij}$ between $\g$, $\e$, and the odd manifold [Fig. 5(c)], where $i$ is the initial state and $j$ is the final state. 
We observed that parity-conserving processes within the even manifold occurred much more quickly than parity-switching processes between the even and odd manifolds. 
Note that the two-fold degeneracy of the odd manifold results in a doubling of the measured rates into the odd states as compared to those out of the odd states.  
Because the two odd states are both decoupled from the resonator, they are indistinguishable in this experiment. 
The hidden Markov model algorithm also yielded the probabilities of the ABS occupying $\g$, $\e$, or the odd manifold at each time step, with the state assignment shown in Fig. 5(b) given by the most likely state. 

\begin{figure}[h]
	\centering
	\includegraphics[width=4.5in]{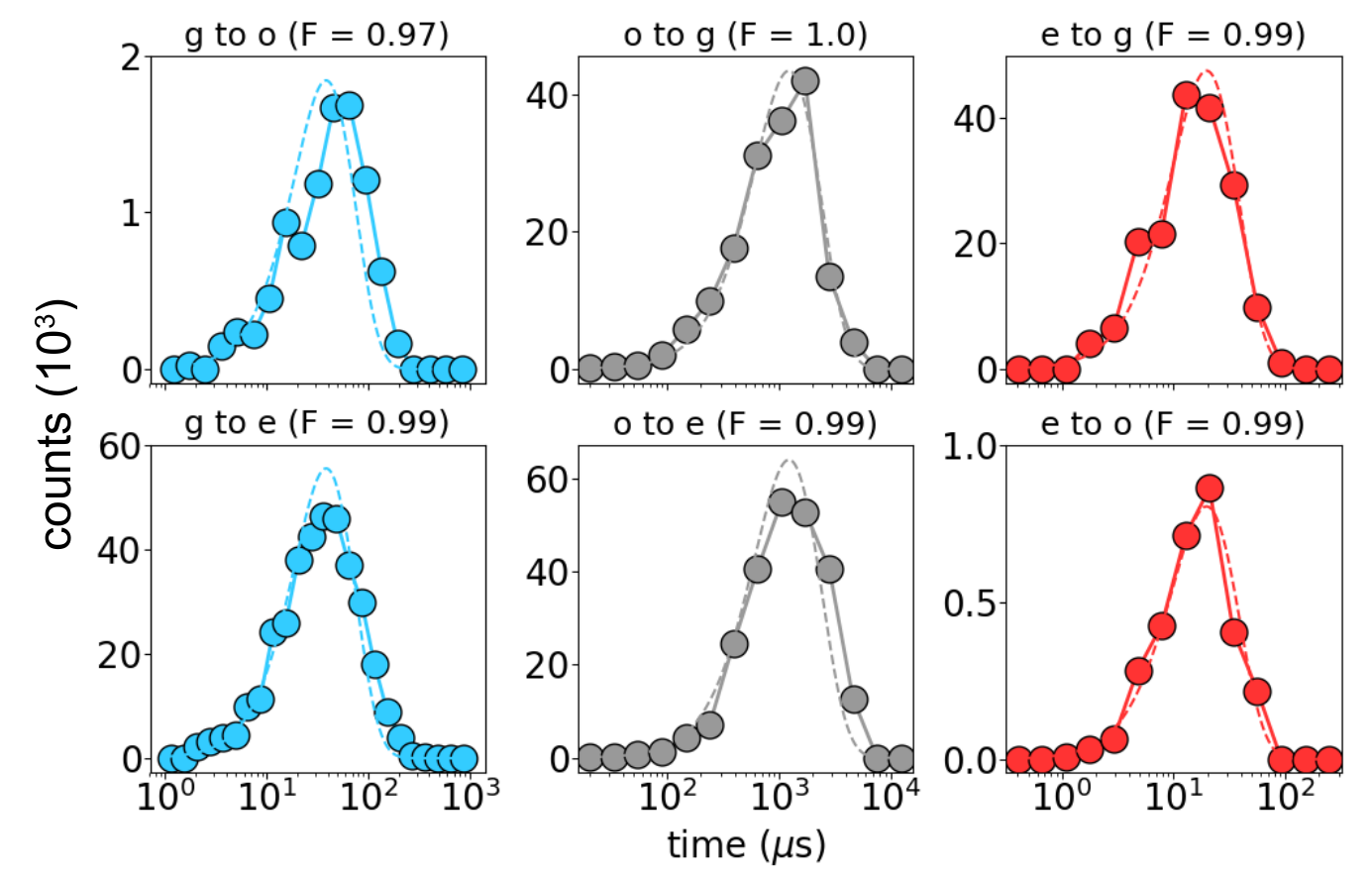}
	\caption{ Histograms of dwell times between jumps of the ABS, weighted by the dwell time. 
	The dashed lines give the predicted distributions assuming the processes are all Poissonian. 
	The fidelity $F$ of the measured histograms to a Poisson process is shown above each plot.  
	}
    \label{rates}
\end{figure}

An assumption of the hidden Markov model analysis is that the underlying processes governing the jumps are Poissonian. 
To verify this, the probability distribution of dwell times between jumps of the system was compared to theory. 
A histogram of the dwell times between any two states should follow an exponential decay $\frac{1}{\bar{\tau}}e^{-\tau/\bar{\tau}}$ where $\tau$ is the dwell time and $\bar{\tau}$ is the average dwell time. 
Following ref. \cite{vool2014}, we instead histogram the dwell times weighted by their length and compare to $\frac{\tau}{\bar{\tau}}e^{-\tau/\bar{\tau}}$ [Fig.~S5] \cite{vool2014}.
This weighting increases the visibility of low-frequency fluctuations of the transition rates.
The fidelity of the data to the theory is computed as

\begin{equation}
	F = \frac{\sum_i \sqrt{M_i P_i}}{\sum_i{M_i}}
\end{equation}

\noindent where $M_i$ are the measured bin values and $P_i$ the predicted. 
The fidelities of all six histograms are 0.97 or above, verifying that the system follows Poisson statistics and that the use of a hidden Markov model is valid. 

We define the parity decay rate $1/\Tp$ as the average rate of population transfer from the odd manifold to the even manifold, plus the average rate of population transfer from the even manifold to the odd manifold. 
Separating the odd manifold into the two states $\ket{o \downarrow}$ and $\ket{o \uparrow}$, the rate from odd to even is given by $\Gamma_\mathrm{odd, \: even} = p_{o \downarrow} (\Gamma_{o \downarrow, g}+\Gamma_{o \downarrow, e}) + p_{o \uparrow} (\Gamma_{o \uparrow, g}+\Gamma_{o \uparrow, e})$ where $p_i$ is the probability for the state to be occupied. 
Because we cannot distinguish between $\ket{o \downarrow}$ and $\ket{o \uparrow}$, we assume $p_{o \downarrow} = p_{o \uparrow} = 0.5$,  $\Gamma_{o \downarrow, g} = \Gamma_{o \uparrow, g} = \Gamma_{o, g}$, and $\Gamma_{o \downarrow, e} = \Gamma_{o \uparrow, e} = \Gamma_{o, e}$. 
With these simplifications, we have $\Gamma_\mathrm{odd, \: even} = \Gamma_{o, g} + \Gamma_{o, e}$. 
Similarly, we assume that $\Gamma_{g, o \downarrow} = \Gamma_{g, o \uparrow} = \Gamma_{g, o}/2$ and $\Gamma_{e, o \downarrow} = \Gamma_{e, o \uparrow} = \Gamma_{e, o}/2$, where the factor of $1/2$ comes from the odd-state degeneracy. This gives $\Gamma_\mathrm{even, \: odd} = p_g\Gamma_{g, o} + p_e\Gamma_{e, o}$, and the final expression for the parity lifetime becomes

\begin{equation}
	\frac{1}{\Tp} = \Gamma_\mathrm{odd, \: even} + \Gamma_\mathrm{even, \: odd} = \Gamma_\mathrm{og}+\Gamma_\mathrm{oe}+p_\mathrm{g}\Gamma_\mathrm{go}+p_\mathrm{e}\Gamma_\mathrm{eo}
\end{equation}

\noindent where the probabilities to be in $\g$ and $\e$ are given by $p_\mathrm{g} = \Gamma_\mathrm{eg}/(\Gamma_\mathrm{ge}+\Gamma_\mathrm{eg})$ and $p_\mathrm{e} = \Gamma_\mathrm{ge}/(\Gamma_\mathrm{ge}+\Gamma_\mathrm{eg})$. 
Plugging in the rates extracted with the hidden Markov model yields $\Tp = 160 \pm 10~\us$.

\end{document}